\documentclass[a4paper,11pt]{JHEP3}
\usepackage{graphicx}
\usepackage{epsfig}
\usepackage{amsmath, amssymb}
\usepackage{dcolumn}% Align table columns on decimal point
\usepackage{bm}% bold math
\usepackage{amsfonts}

\usepackage{subfigure}
\usepackage{amssymb}
\usepackage{epstopdf}

\newcommand{\beq}{\begin{equation}}
\newcommand{\eeq}{\end{equation}}
\newcommand{\bea}{\begin{eqnarray}}
\newcommand{\eea}{\end{eqnarray}}

\newcommand{\bi}{\begin{itemize}}
\newcommand{\ei}{\end{itemize}}

\newcommand{\ps}{(2\pi)^3}
\newcommand{\IR}{{_{\rm IR}}}
\newcommand{\UV}{{_{\rm UV}}}

\newcommand{\hI}{\hspace{1cm}}

\newcommand{\hV}{\hspace{.5cm}}

\newcommand{\lb}{\left\lbrace}
\newcommand{\rb}{\right\rbrace}
\newcommand{\nn}{\nonumber}

\newcommand{\bk}{{\mathbf k}}

\newcommand{\bx}{{\mathbf x}}

\newcommand{\bq}{{\mathbf q}}

\author{Kari Enqvist, Daniel G. Figueroa\\
University of Helsinki and Helsinki Institute of Physics, P.O.Box 64,
FIN-00014 University of Helsinki, Finland}
\author{Gerasimos Rigopoulos\\
Institute for Theoretical Particle Physics and Cosmology, RWTH Aachen, D - 52056, Germany
}

\keywords{MSSM flat-directions, de Sitter Inflation, Langevin Equations}

\preprint{HIP-2011-24/TH,  TTK-11-39}

\title{Fluctuations along supersymmetric flat directions during inflation}

\abstract{We consider a set of scalar fields, consisting of a single flat direction and one or several non-flat directions. We take our cue from the MSSM, considering separately D-flat and F-flat directions, but our results apply to any supersymmetric scenario containing flat directions. We study the field fluctuations during pure de Sitter inflation, following the evolution of the infrared modes by numerically solving the appropriate Langevin equations. We demonstrate that for the Standard Model $U(1)_{\rm Y}$, $SU(2)_{\rm L}$ or $SU(3)_{\rm c}$
gauge couplings, as well as for large enough Yukawa couplings, the fluctuations along the non-flat directions effectively block the fluctuations along the flat directions. The usual expected behaviour $\langle \phi^2\rangle \propto N$, with $N$ the number of efolds, may be strongly violated, depending on the coupling
strengths. As a consequence, those cosmological considerations, which are derived assuming that during inflation flat directions fluctuate freely, should be revised.
}

\begin{document}
\section{Introduction}
\label{sec:Introduction}

The scalar potential of a supersymmetric gauge theory has always a number of flat directions, along which the potential vanishes identically. In particular, the scalar field space of the MSSM (Minimal Supersymmetric Standard Model) is 49-dimensional, and there are some three hundred flat directions; these have all been classified (for a review, see \cite{MSSM-REV}). The flat directions are rays in field space, restricted to lie in some subspaces. More concretely, flat directions correspond to  configurations where some of the field values are related to each other while the rest are set to zero. One consequence is that most of the flat directions are not simultaneously flat; if there happens a field fluctuation along some given flat direction, many other potentially flat directions will no longer remain flat. This is true in general for any supersymmetric gauge theory with scalars.

In the absence of supersymmetry breaking, flat directions are perturbatively safe as they are protected by a non-renormalization theorem~\cite{Non-RenonTh}. However, they can be (and are) lifted by supersymmetric non-renormalizable terms. For each flat direction, the lowest order non-renormalizable operators lifting the flatness are known and classified. Supersymmetry breaking will also induce non-flatness through the mass terms and A-terms, but for the purpose of the present paper they are not important.

For concreteness, we will focus on the MSSM, although everything that we have to say will generically hold for any supersymmetric model.
A central ingredient in the cosmological considerations of flat directions is often the tacit assumption that during inflation, fields can fluctuate along a given flat direction essentially like massless fields in de Sitter space. The variance of a massless field $\phi$ in pure de Sitter space is given by \cite{bunchdavies}
\beq\label{oldresult}
\langle\phi^2\rangle =\frac{1}{4\pi^2}H_o^2N,
\eeq
where $N$ is the number of e-folds and $H_o$ is the Hubble parameter during inflation. Fluctuations are assumed to continue growing in amplitude until the non-renormalizable terms take over, making the effective mass $V'' > H_o^2$, and hence stopping the mean-square-root-amplitude from evolving any further. Quantitatively, for a flat direction with mass $m\ll H_o$, the asymptotic behaviour $t\to\infty$ of the variance is given by~\cite{eno}
\beq\label{eno}
\langle\phi^2\rangle = \frac{3H_o^4}{8\pi^2m^2}\left(1-\exp\left(-\frac{2m^2t}{3H_o}\right)\right)~,
\eeq
Hence if $m\sim H_o$, the mean-root-square amplitude of the fluctuations would saturate to a constant value. %decays exponentially.
Assuming that the effective mass is only due to the %kicks in only because of a
non-renormalizable term(s), one would obtain a very large flat direction field amplitude that depends on the dimension of the non-renormalizable term(s). Taking $V\sim \phi^6/M_p^2$ as an example, with $M_p \approx 2.3\cdot10^{18}$ GeV the reduced Planck mass, once the condition $V'' \sim H_o^2$ is attained, one would find a final amplitude of $\phi_\infty\sim (H_oM_p)^{1/2}\gg H_o$. Note that this additionally assumes that inflation lasts long enough as reaching the asymptotia in this particular case would require $N\simeq M_p/H_o$ efolds. However, this is not quite true since from (\ref{eno}) we see that equilibrium is attained
when $V\sim H_o^4$, corresponding to $\phi_{\rm eq}\sim (H_o^4M_p^2)^{1/6}=(H_o/M_p)^{1/6}\phi_\infty$.
The equilibrium probability distribution $\mathcal{P}$ is not peaked at $\phi_{\rm eq}$ but is flat with $\mathcal{P}\sim \exp(-V/H_o^4)$~\cite{starobinsky}.

However, during inflation all fields fluctuate, including the non-flat directions. Since the existence of a given flat direction is conditional upon most of the fields staying at the origin, one may wonder how likely it is that a large amplitude along a flat direction can be obtained, considering that the flat direction is actually coupled to non-flat directions, which are also fluctuating. They may in effect provide an effective mass for the flat directions, thus preventing the spreading of the flat direction variances. We will demonstrate that in most cases, depending on the gauge and Yukawa coupling strengths, the small fluctuations of the non-flat directions are indeed sufficient to block the growth of the mean-root-square amplitude along the flat direction.

In the present paper we address this issue by separating the ultraviolet (UV) and infrared (IR) field modes and solving the appropriate infrared Langevin equations for a restricted set of MSSM fields in de Sitter space. We consider the components of the Higgs fields $H_u$ and a slepton field $L$ that are mostly non-flat field directions but, in a certain combination, also include a flat direction;
we also include other non-flat field directions. Our conclusions will be, in any case, extrapolable to other scenarios with flat and non-flat directions interacting.

The contents of the paper are as follows. After the introduction presented here in Section~\ref{sec:Introduction}, the fields and the potential are written out in Section~\ref{sec:FlatAndNonFlatDirectionsPotential}. In Section~\ref{sec:StochasticDynamics} we review the formalism of the Langevin equations and the separation of the IR and UV modes, and describe our procedure for the numerical implementation. In Section~\ref{sec:ResultsAndDiscussions} we present the outcome of the numerical solutions we obtain, and discuss the role of the coupling strengths. We treat F-flatness and D-flatness separately. In Section~\ref{sec:Conclusions} we draw our conclusions.

\section{Coupling of flat and non-flat directions}
\label{sec:FlatAndNonFlatDirectionsPotential}

\subsection{The scalar potential}
A supersymmetric potential is given by
the sum of the F-term and the D-term:
\beq
V_S\equiv V_F+V_D=\sum_i\left|\frac{df}{d\phi_i}\right|^2+\frac 12 \sum_ag_a^2D^aD^a~,
\eeq
where $\phi_i$ stands for all the scalar fields. In the MSSM the superpotential reads
\beq
f=\lambda_uQH_u\bar u+\lambda_dQH_d\bar d+\lambda_eLH_d\bar e+\mu H_uH_d~,
\eeq
where $Q$ and $L$ are respectively the squark and slepton doublets, and $H_u$ and $H_d$ are the Higgses.
Here $\lbrace g_a \rbrace$ are the Standard Model gauge couplings, $\lbrace \lambda_i \rbrace$ the Yukawas, and
\beq
D^a=\sum\phi_i^\dagger T^a\phi_i
\eeq
is the D-term. If for some set of field values $F=0$ ($D=0$), we call the corresponding direction in field space F-flat (D-flat).

During inflation, the fields will typically obtain an induced mass term with $m\sim H_o$. However, the magnitude and also the existence
of such a term depends on the details of the inflationary sector (for a review, see e.g. \cite{MSSM-REV}). For instance, in D-term inflation
the Hubble-induced mass term is absent. In what follows we will ignore it, which in the light of our results is the conservative
assumption since the Hubble-induced mass will usually only add to the blocking of fluctuations (although it is also possible
that the Hubble-induced mass term has a negative sign).

To be concrete, we will focus on the leptonic sector of MSSM only and take our cue from the simple $LH_u$ flat direction,
given by
\beq\label{HLflat}
H_u=\left(\begin{array}{c}0\\ \phi\end{array}\right),~
L=\left(\begin{array}{c}\phi\\ 0\end{array}\right)
\eeq
with all the other scalar fields $=0$. It is easy to verify that the configuration (\ref{HLflat}) is both F-flat and D-flat.
Thus we will assume that the squark fields do not fluctuate; since they are not coupled to the leptonic sector,
their fluctuations would affect only the Higgses (but nevertheless also the $LH_u$ flat direction). In addition, we also set $H_u^+=0$
for simplicity.

After an $SU(2)$ rotation,
the remaining degrees of freedom are:
\beq
\label{thefields}
H_d=\left(\begin{array}{c}H^0\\ 0\end{array}\right),~
L=\left(\begin{array}{c}\nu\\ e\end{array}\right),~~\bar e~,~~\phi,
\eeq
where we have denoted the prospective flat direction as $H_u^0\equiv\phi$. As we will see, some of these dofs are, qualitatively speaking, dynamically redundant.

Given
the fields (\ref{thefields}), we find
\beq\label{Fterms}
V_F=\lambda_e^2\left(|eH^0|^2+|e\bar e|^2+|\bar eH^0|^2+|\nu\bar e|^2\right)~,
\eeq
while
\bea\label{Dterms}
D_2&=&\frac 12\left(|\nu|^2-|e|^2+|H^0|^2-|\phi|^2\right)~,\nonumber\\
D_1&=&\frac 12\left(-|\nu|^2-|e|^2-|H^0|^2+2|\bar e|^2+|\phi|^2\right)
\eea
are respectively the $SU(2)_L$ and $U(1)_Y$ D-terms. As is conventional, we neglect the $\mu$-term which is of the
order of susy breaking mass.

From (\ref{thefields}) and (\ref{Dterms}) we readily observe that the flat direction is given by $\nu=\phi$, provided
the right-handed slepton $\bar e = 0$.

\subsection{Fluctuations breaking F-flatness}

Note that in general the flat direction $\phi$ mixes with the non-flat part of $\nu$
in the kinetic term (a similar mixing
arises also through susy breaking mass terms).  This is just an artifact of the chosen notation which presents a computational
nuisance, since in the evolution equations for fluctuations, one would have to diagonalize the kinetic term at each time step. However, this
can be avoided if we choose $\phi$ to be real (with $\phi\to \phi/\sqrt{2}$) and write
\beq\label{nuimag}
\nu=\frac {1}{\sqrt{2}}\left(\phi+ih\right)~,
\eeq
where $h$ is real and represents the difference $\nu-H_u^0$. Then the kinetic terms of all fields will remain diagonal.
We do not believe that the restriction (\ref{nuimag}), which represents a subclass of all possible fluctuations,
changes the outcome of the dynamics of the F-term in any qualitative way.
However,  in the D-term the choice (\ref{nuimag}) decouples the flat direction $\phi$ from the non-flat directions completely.
Therefore the effect of the fluctuations of non-flat directions on the flat direction in the D-term must be considered separately
and will be addressed below.

Let us now inspect (\ref{Fterms}) and (\ref{Dterms}) to see what sort of generic couplings we obtain. Let us also denote
the non-flat directions generically by $\chi_i$. We see that from the F-term one obtains potential terms of the type $|\chi_i\chi_j|^2$
with $i\ne j$, as well as a coupling of the flat direction to non-flat direction $\bar e$ through $|\nu\bar e|^2=(\phi^2+h^2)|\bar e|^2$.
With the form (\ref{nuimag}), there is no $\phi$-dependence in the D-terms, and one obtains
potential terms of the form $|\chi_i\chi_j|^2$ with $i\ne j$, as well as terms like $|\chi_i|^4$. Therefore, the generic form of the
potential when some non-flat excitations are coupled to the flat direction in the F-term, can be summarized by the example
\beq\label{simulatethis}
V_S=\frac 12\lambda_e^2(\phi^2+h^2)\bar e^2+\frac 18g_2^2h^4+\frac 18g_1^2(h^4+4\bar e^4-4h^2\bar e^2)+\frac{\phi^6}{M_p^2}~,
\eeq
where we have chosen the fields to be real (and properly normalized) and have set $e=H^0=0$ since their contribution to the potential is qualitatively similar
to $\bar e$; as such, they just represent some additional fluctuating dofs that couple to each other but not to the flat direction.
The last term in (\ref{simulatethis}) is the non-renormalizable term appropriate to the $LH_u$ flat direction, which is known to be
lifted by $d=4$ operator in the superpotential. For other flat directions
the dimension can be different but always $d\ge 4$; all flat directions are lifted by  operators with $d\le 9$  \cite{Dine:1995kz,Gherghetta:1995dv}. It will turn out that,
for most cases, the non-renormalizable term
 is irrelevant for determining the final amplitude along the flat direction.

We will treat a potential of the type (\ref{simulatethis}) as a generic example describing a flat direction $\phi$ and two non-flat directions that may all fluctuate during inflation;
however, as noted above, the full spectrum of fluctuations is not included as the flat and non-flat directions remain decoupled in the D-term.
In what follows, we assume that the Hubble rate during inflation is much larger than the susy breaking scale. Hence we will neglect both the supersymmetry breaking mass terms and A-terms in the potential.

\subsection{Fluctuations breaking D-flatness}

To study the effect of fluctuations inside the D-term, it is not convenient to consider the flat direction as a background solution.
Rather, we simply decouple the F- and D-terms by setting the Yukawas to zero and consider the field fluctuations in the potential
\beq\label{dtermcase}
V_D=\frac{1}{8}g^2(L^2-H^2)^2~,
\eeq
where $L$ and $H$ have been assumed to be real dofs for simplicity. Their kinetic terms are diagonal. The flat direction is the ray $L=H$, and we will consider the evolution of the probability
distribution due to inflationary fluctuations in the $(H,L)$ plane. The fluctuations of other (non-flat) fields are ignored for simplicity. Although (\ref{dtermcase}) is motivated by
MSSM, we will treat the coupling $g$ as essentially a free parameter, focusing however on values that are appropriate for the MSSM.\\

%%%%%%%%%%%%%%%%%%%%%%%%%%%%%%%%%%%%%%%%%%%%%%%%%%%%%%%%%%%%%
\section{Stochastic dynamics of fields during inflation.}
\label{sec:StochasticDynamics}
%%%%%%%%%%%%%%%%%%%%%%%%%%%%%%%%%%%%%%%%%%%%%%%%%%%%%%%%%%%%%

Let us now study the field fluctuations during inflation, assuming a constant Hubble rate $H_o$. Any field in de Sitter (or quasi-de Sitter) can be decomposed into IR modes (with momenta $k < \epsilon aH_o$) and UV modes (with $k > \epsilon aH_o$), where $\epsilon$ is a constant smaller than unity, $\epsilon<1$. The IR modes have non-trivial stochastic dynamics~\cite{Starobinsky86} due to the continuous influx of UV modes around the Hubble radius, which can be considered as classical stochastic fluctuations as long as $\epsilon$ is sufficiently small; any value $\epsilon \lesssim 1$ will do. In this manner the UV modes impart continuous random 'kicks' on the IR sector that are superimposed over the usual deterministic evolution. The dynamics are described by multi-field Langevin type equations, from which one can extract the probability distribution function (PDF) of the fields' fluctuations.
%%%%%%%%%%%%%%%%%%%%%%%%%%%%%%%%%%%%%%%%%%%%%%%%%%%%%%%%%%%%%%%%%%%%%%%%%%%%%%%%%%%%%%%%%%%%%%%%%%%%%%%%%%%%
%%%%%%%%%%%%%%%%%%%%%%%%%%%%%%%%%%%%%%%%%%%%%%%%%%%%%%%%%%%%%%%%%%%%%%%%%%%%%%%%%%%%%%%%%%%%%%%%%%%%%%%%%%%%
\subsection{The Langevin equations}

We begin by briefly reviewing the formalism of the IR stochastic dynamics during inflation~\cite{Starobinsky86} (see also eg~\cite{Nakao:1988yi, Stewart91} for some early references on the subject).
Let us define the IR and UV parts of a scalar field $\phi(\bx,t)$ as
\begin{eqnarray}\label{eq:IR/UV}
\phi_\IR(\bx,t) &=& \int \frac{d^3\bk}{\ps}\,e^{-i\bk\bx}\,\phi_\bk(t)\,W(\bk,t) \\
\label{eq:IR/UVbis}
\phi_\UV(\bx,t) &=& \int \frac{d^3\bk}{\ps}\,e^{-i\bk\bx}\,\phi_\bk(t)\,[1-W(\bk,t)]\,,
\end{eqnarray}
where the function $W(\bk,t)$ is an IR-filter (window function) that is subject to constraints
\begin{eqnarray}
 W(\bk,t) \xrightarrow{|\bk| \ll Q(t)} 1\,,\hI W(\bk,t) \xrightarrow{|\bk| \gg Q(t)} 0\,,\hI Q\left.\frac{dW}{dk}\right|_{k \sim Q} \ll -1\,,
\end{eqnarray}
with $Q(t)$ a time-dependent scale. For inflation, the relevant characteristic scale is the (comoving) Hubble radius, $\mathcal{H} = aH_o$, which provides a natural border between the UV and IR dynamics of the modes of a light scalar field. We will thus identify $Q(t) = \epsilon \mathcal{H}$, with $\epsilon < 1$. Since the energy density of the fields will be sub-dominant they will not back-react on the metric during inflation, and we will ignore gravitational perturbations; for their inclusion in a stochastic formalism see \cite{Salopek:1990re, Rigopoulos:2004gr}.

Consider now a set of interacting fields in de Sitter space, $\lbrace \phi^i\rbrace$. Each field can be decomposed as $\phi_i(\bx,t) = \Phi_i(\bx,t) + \varphi_i(\bx,t)$, with $\Phi_i$ and $\phi_i$ the IR and UV parts obtained according to eqs.~(\ref{eq:IR/UV})-(\ref{eq:IR/UVbis}). In order to most accurately follow the dynamics of the IR dof, $\Phi_i$ and $\dot\Phi_i$ should be considered as independent variables (though linked through the $eom$) and a Hamiltonian formulation is the most natural framework to use. Calling $\pi_i$ the conjugated momentum of $\phi_i$, the $eom$ in the hamiltonian picture are
\begin{eqnarray}\label{eq:HamiltonianEqs}
\dot\phi_i = \pi_i\,,\hI\hV \dot\pi_i + 3H_o\pi_i &=& \frac{1}{a^2}\nabla^2\phi_i - d_iV\,,
\end{eqnarray}
with $d_iV \equiv \partial V/\partial\phi_i$. We then IR/UV decompose $\phi_i$ and $\pi_i$ independently as $\phi_i(\bx,\eta) = \Phi_i(\bx,\eta) + \varphi_i(\bx,\eta)$ and $\pi_i(\bx,\eta) = \Pi_i(\bx,\eta) + \delta\pi_i(\bx,\eta)$, and introduce such a decomposition into
eqs.~(\ref{eq:HamiltonianEqs}). We provide more details in the Appendix. One finally finds that the dynamical equations for the IR dofs are
\begin{eqnarray}\label{eq:LangevinEqs bis}
&&  \dot\Phi_i = \Pi_i + s^{(\phi)}_i(\bx,\eta)\,,\\
\label{eq:LangevinEqs bis2}
&& \dot\Pi_i = -3H_o\Pi_i -D_i\bar V + s^{(\pi)}_i(\bx,\eta)\,,
\end{eqnarray}
with $D_i = \partial/\partial\Phi_i$, $\bar V$ the potential taken as a function only of the IR components, i.e. $\bar V \equiv V(\lb \Phi_j\rb)$, and
\begin{eqnarray}\label{noise1}
s^{(\phi)}_i(\bx,t) \equiv \int{\frac{d^3\bk}{\ps}\,e^{-i\bk\bx}\,\phi_i(\bk,t)\,\dot W(k,t)}%\int{\frac{d^3\bk}{\ps}\,\dot W(\bk,t)\left(\hat a_k e^{-i\bk\bx}f_{\phi_i}(\bk,t) + \hat a_k^\dag e^{+i\bk\bx}f_{\phi_i}^*(\bk,t)\right)}
\\\label{noise2}
s^{(\pi)}_i(\bx,t) \equiv \int{\frac{d^3\bk}{\ps}\,e^{-i\bk\bx}\,\pi_i(\bk,t)\,\dot W(k,t)}\,.
%\int{\frac{d^3\bk}{\ps}\,\dot W(\bk,t)\left(\hat a_k e^{-i\bk\bx}f_{\pi_i}(\bk,t) + \hat a_k^\dag e^{+i\bk\bx}f_{\pi_i}^*(\bk,t)\right)}\,,
\end{eqnarray}
Note that since we are interested in the leading order IR behavior we have dropped the gradient terms from (\ref{eq:LangevinEqs bis}) and (\ref{eq:LangevinEqs bis2}), and have also ignored any corrections coming from the IR/UV decomposition of an interacting potential $V$.

Eqs.~(\ref{eq:LangevinEqs bis}) and~(\ref{eq:LangevinEqs bis2}) are operator equations. The terms $e^{-i\bk\bx}\phi_k$, $e^{-i\bk\bx}\pi_k$ should be understood as $\hat a_ke^{-i\bk\bx}f_\phi(\bk)$ + $\hat a_k^\dag e^{+i\bk\bx}f_\phi^*(\bk)$, $\hat a_k e^{-i\bk\bx}f_\pi(\bk)$ + $\hat a_k^\dag e^{+i\bk\bx}f_\pi^*(\bk)$, respectively, with $a_k, a_k^\dag$ the usual creation/annihilation operators, and $f_\phi(\bk),f_\pi(\bk)$ the field mode functions. However, if the window function is chosen appropriately such that the UV/IR split occurs shortly after the relevant modes have crossed the horizon, the phase of the mode functions entering (\ref{noise1}) and (\ref{noise2}) becomes almost constant and then $s^{(\phi)}_i$ and $s^{(\pi)}_i$ commute with each other at different times. They can be considered as classical stochastic forces and the equations for the IR dofs are therefore Langevin-type equations. Ignoring the self-interactions of the UV modes makes the stochastic terms gaussian random fields so that all the statistical information is encoded in the correlators
\begin{eqnarray}
s^{(\phi)}_{ij}(t,t') \equiv \left\langle s^{(\phi)}_i(\bx,t)s^{(\phi)}_j(\bx,t') \right\rangle \,,\\
s^{(\pi)}_{ij}(t,t') \equiv \left\langle s^{(\pi)}_i(\bx,t)s^{(\pi)}_j(\bx,t') \right\rangle\,.
\end{eqnarray}
Note that since we are not interested in the spatial correlations, we are only considering the correlators at the same spatial point, which in reality corresponds to a region of physical volume $V\sim 1/H_o^3$. Points separated by physical distances $L>1/H_o$ are essentially uncorrelated.

Apart from the classical evolution dictated by the deterministic parts of (\ref{eq:LangevinEqs bis}) and (\ref{eq:LangevinEqs bis2}), the stochastic forces acting over a small time interval $\delta t \ll 1/H_o$ displace the fields by \begin{eqnarray}
\delta\phi_i &=& \int_t^{t+\delta t}s^{(\phi)}_i(t')dt'\,,\\
\delta\pi_i &=& \int_t^{t+\delta t}s^{(\pi)}_i(t')dt'\,.
\end{eqnarray}
Therefore, the correlators we really need are
\begin{eqnarray}
&& S^{(\phi)}_{ij}(t,t' ; \delta t) \equiv \int_{t}^{t+\delta t}\int_{t'}^{t'+\delta t}s^{(\phi)}_{ij}(\tau,\tau')\,d\tau d\tau'\,,\\
&& S^{(\pi)}_{ij}(t,t' ; \delta t) \equiv \int_{t}^{t+\delta t}\int_{t'}^{t'+\delta t}s^{(\pi)}_{ij}(\tau,\tau')\,d\tau d\tau'\,.
\end{eqnarray}
In  pure inflationary de Sitter background, the solution to the mode equations (with boundary conditions matching Minkowski modes at $k \rightarrow \infty$) are well known, see  for instance~\cite{LindeBook}. Using such mode functions, $f_{\phi_i}(\bk,t)$ and $f_{\pi_i}(\bk,t)$, evaluated at $k = \epsilon aH_o$, and choosing a step-function for the IR-filter, $W(k,t) = \theta( \epsilon aH_o-k)$, the equal-time correlators we need are found\footnote{See the Appendix for more details.} to be
\begin{eqnarray}\label{eq:CorrelatorsNaturalVariables}
S^{(\phi)}_{ij}(t,dt) = \delta_{ij}\,(1+\epsilon^3)\frac{H_o^3}{4\pi^2}\,dt\,,\hI
S^{(\pi)}_{ij}(t,dt) = \delta_{ij}\,\epsilon^4\,\frac{H_o^5}{(2\pi)^2}\,dt\,,
\end{eqnarray}
valid only in the limit $dt \ll 1/H_o$. The choice $\epsilon < 1$ then ensures that the classical interpretation of the stochastic noise terms is valid. We also see that for $\epsilon<1$ the momentum noise term is suppressed and the main stochastic component lies in the fluctuations of $\phi$.

%%%%%%%%%%%%%%%%%%%%%%%%%%%%%%%%%%%%%%%%%%%%%%%%%%%%%%%%%%%%%%%%%%%%%%%%%%%%%%%%%%%%%%%%%%%%%%%%%%%%%%%%%%%%
%%%%%%%%%%%%%%%%%%%%%%%%%%%%%%%%%%%%%%%%%%%%%%%%%%%%%%%%%%%%%%%%%%%%%%%%%%%%%%%%%%%%%%%%%%%%%%%%%%%%%%%%%%%%
\subsection{Numerical implementation}

\noindent Rescaling the field variables as
\begin{eqnarray}
\begin{array}{rclcrcl}
\Phi_i &\rightarrow& \tilde\Phi_i = \Phi_i/H_o & \hI,\hI& \hI s^{(\phi)} &\rightarrow& \tilde s^{(\phi)} =  s^{(\phi)}/H_o^2\vspace{1mm}\\
\Pi_i &\rightarrow& \tilde\Pi_i = \Pi_i/H_o^2 & \hI,\hI& \hI s^{(\pi)} &\rightarrow& \tilde s^{(\pi)} =  s^{(\pi)}/H_o^3\\
\end{array}\\
\bar V(\lb\Phi_j\rb) \rightarrow \tilde{\bar V}(\lbrace\tilde\Phi_j\rbrace)/H_o^4\,,\hI\hI\hV\,
\end{eqnarray}
makes it possible to express the Langevin Eqs.~as depending only on dimensionless variables,
\begin{eqnarray}
 \tilde\Phi_i'(N) &=&  \tilde\Pi_i(N) + \tilde s_i^{(\phi)}\,,\\
 \tilde\Pi_i'(N) &=& -3H_o\tilde\Pi_i(N) - D_i\tilde{\bar V}(\lb\Phi_j(N)\rb) + \tilde s_i^{(\pi)}\,,
\end{eqnarray}
with $'$ standing for derivatives with respect the number of e-folds $N = \int Hdt$.

In order to solve these equations in a computer, we just need to discretize them by choosing a small time step, i.e.~$dN \ll 1$. Dropping the tildes for clarity of the notation, one arrives at the discretized (dimensionless) iterative equations
\begin{eqnarray}\label{eq:DiscreteLangevinEqs}
 \Phi_i(N+dN) &=& \Phi_i(N) + \Pi_i(N)dN + S_i^{(\phi)}\\
 \Pi_i(N+dN) &=& \Pi_i(N) - 3\Pi_i(N)dN - D_i\bar V(\lb\Phi_j(N)\rb)dN + S_i^{(\pi)}\,,
\end{eqnarray}
where the stochastic terms are drawn from a gaussian random distribution with correlators
\begin{eqnarray}\label{eq:CorrelatorsDimensionlessVariables}
S^{(\phi)}_{ij}(dN) &\equiv& \langle S^{(\phi)}_iS^{(\phi)}_j \rangle = \delta_{ij}\,(1+\epsilon^3)\frac{dN}{4\pi^2}\,,\\
\label{eq:CorrelatorsDimensionlessVariables2}
S^{(\pi)}_{ij}(dN) &\equiv& \langle S^{(\pi)}_iS^{(\pi)}_j \rangle = \delta_{ij}\,\epsilon^4\,\frac{dN}{4\pi^2}\,.
\end{eqnarray}

Eqs.~(\ref{eq:DiscreteLangevinEqs})-(\ref{eq:CorrelatorsDimensionlessVariables2}) characterize completely the dynamics of the IR dof of any set of scalar fields during de~Sitter inflation. Note that the only scale of the problem\,\footnote{Of course, non-renormalizable terms in the potential will also introduce new scales, but as said before, and as we demonstrate with the numerics, such scales never play a role in the problem under study.}, the inflationary Hubble constant $H_o$, has been scaled out, so Eqs.~(\ref{eq:DiscreteLangevinEqs})-(\ref{eq:CorrelatorsDimensionlessVariables2}) are indeed scale-free equations. Any dimension-full functional built by powers of fields, for instance $\phi^2$, will then be measuring an amplitude in units corresponding to the same powers of $H_o$. Eqs.~(\ref{eq:DiscreteLangevinEqs})-(\ref{eq:CorrelatorsDimensionlessVariables2}) are therefore universal in this sense, since the physics they describe is independent of the inflationary scale. It is also worth mentioning that the stochastic character is time-independent, since the correlators $S^{(X)}_{ij}$ do not depend on the 'time' $N$, but only on the step $dN$. The dynamical behaviour of the IR dof should, of course, not depend on such step $dN$. Thus, after solving the system for a given step $dN \ll 1$, one should always make sure that the same dynamics is recovered by choosing, for instance, $dN/10$. If that is not the case, one must then decrease further the step $dN$, until finding that the fields' statistical properties are insensitive to further decrements. Checks of this nature have been performed on all the numerical results we present in the next section.

\FIGURE[ht]{
\epsfig{width=11.0cm,height=7.0cm,angle=0, file=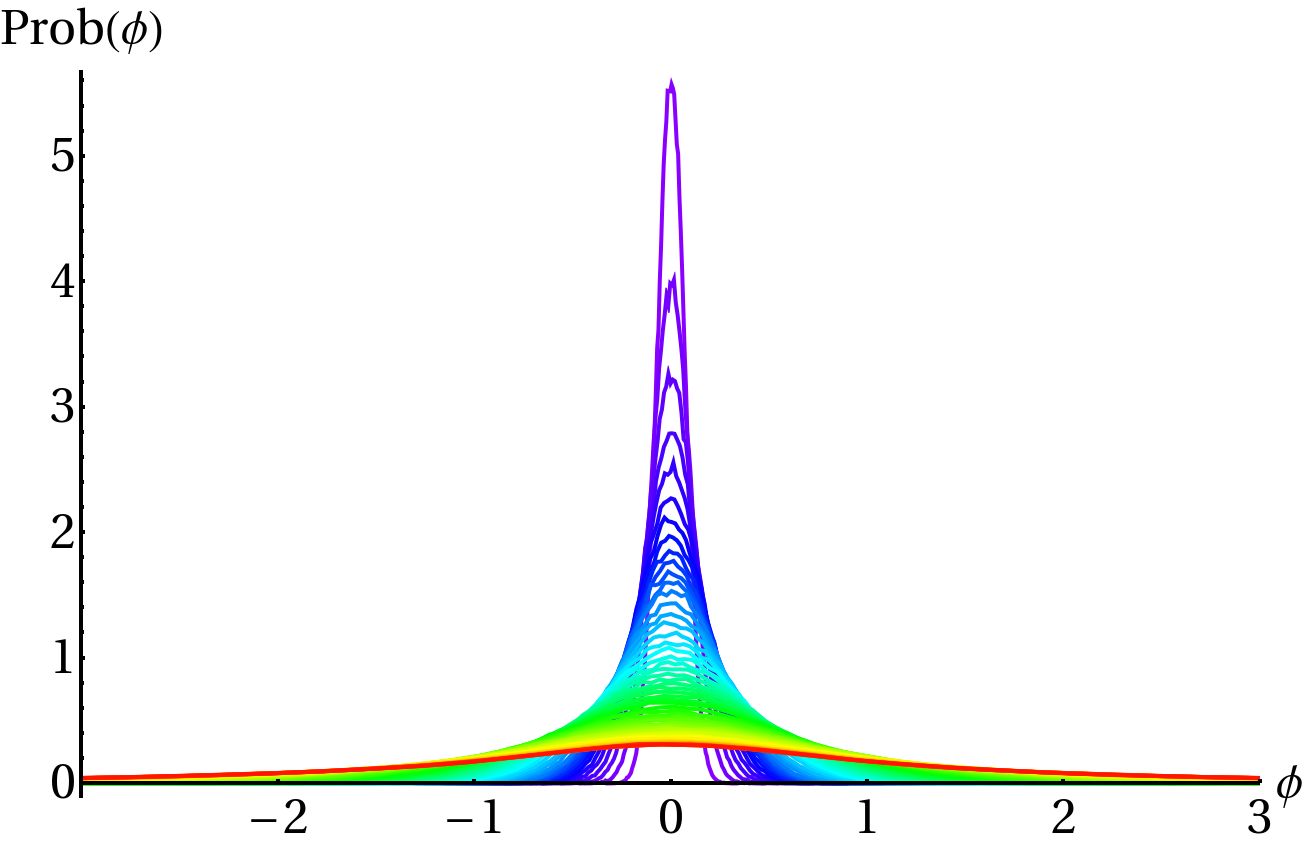}
\caption{Time evolution of the Probability distribution function $\mathcal{P}(\phi)$ of the Flat direction $\phi$. The PDF at each time/e-fold $N$ was reconstructed with $10^5$ independent runs, all for the case $\lambda = g_1 = g_2 =1$. Each PDF shown in the sequence was obtained by dividing the unit-normalized histogrammatic distributions of $\phi$ by its field bin width at each time. The colors represent different stages of the evolution. Purple-to-DarkBlue, from $N = 0.1$ to $N = 1$ e-folds; %, in intervals of $\Delta N = 0.1$.
DarkBlue-to-LightBlue, from $N = 1$ to $N = 3.5$ e-folds; % in intervals of $\Delta N = 0.2$.
Green, from $N = 3.5$ to $N = 10$ e-folds; %, in intervals of $\Delta N = 0.5$.
Green-to-yellow, from $N = 10$ to $N = 100$ e-folds; %, in intervals of $\Delta N = 5$.
and Yellow-to-Red, from $N = 100$ to $N = 1500$ e-folds.%, in intervals of $\Delta N = 20$.
}
\label{fig:FlatDirectionPDF}
}
%%%%%%%%%%%%%%%%%%%%%%%%%%%%%%%%%%%%%%%%%%%%%%%%%%%%%%%%%%%%%%%%%%%%%%%%%%%%%%%%%%%%%%%%%%%%%%%%%%%%%%%%%%%%
%%%%%%%%%%%%%%%%%%%%%%%%%%%%%%%%%%%%%%%%%%%%%%%%%%%%%%%%%%%%%%%%%%%%%%%%%%%%%%%%%%%%%%%%%%%%%%%%%%%%%%%%%%%%
\section{Numerical results for fluctuations}
\label{sec:ResultsAndDiscussions}
%%%%%%%%%%%%%%%%%%%%%%%%%%%%%%%%%%%%%%%%%%%%%%%%%%%%%%%%%%%%%%%%%%%%%%%%%%%%%%%%%%%%%%%%%%%%%%%%%%%%%%%%%%%%
%%%%%%%%%%%%%%%%%%%%%%%%%%%%%%%%%%%%%%%%%%%%%%%%%%%%%%%%%%%%%%%%%%%%%%%%%%%%%%%%%%%%%%%%%%%%%%%%%%%%%%%%%%%%
\subsection{Breaking F-flatness}

We have solved the previously discretized equations for the potential~(\ref{simulatethis}). In what follows, we will denote the flat direction as $\phi$ while $\chi_1, \chi_2$ are the non-flat directions. We have chosen for the initial conditions $\phi = \chi_1 = \chi_2 = 0$ at $N = 0$. For a given choice of the couplings $\lbrace \lambda, g_1, g_2 \rbrace$, we have made $n_r$ ($\gg 1$) independent runs with the same initial conditions. Thus each run represents a different realization of the time evolution of the fields' fluctuations. Collecting at a given moment the amplitude of the fields from all the independent runs, we can then reconstruct the probability distribution function (PDF) $\mathcal{P}(\varphi)$ of any field $\varphi$ %(= $\phi, \chi_1, \chi_2$)
at such moment. By definition $\mathcal{P}(\varphi')d\varphi$ represents the probability of the field $\varphi$ to have an amplitude within $[\varphi',\varphi'+d\varphi]$. Thus $\mathcal{P}(\varphi)$ can be obtained from the numerics by dividing the (unit-normalized) histogrammatic distribution of $\varphi$ by its field bin width. Doing this at successive moments, we can track the time evolution of the PDF of each field.

If there was no coupling between the flat and the non-flat directions, $\phi$ would freely fluctuate with an increasing variance $\langle \phi^2\rangle \sim NH_o^2$, until the non-renormalizable term $\phi^6/M_p^2$ would become important and equilibrium would be reached. The non-flat directions also fluctuate, obtaining variances of the order $\sim 0.1 H_o^2$ in a matter of few efolds (bottom lines in Figure~\ref{fig:FlatDirectionVariance}). Because of the coupling of $\phi$ to the fluctuations of non-flat directions $\chi_i$, the flat direction can then obtain an effective mass of order $\sim 0.1 \lambda H_o$. Unless the couplings are very small, $\phi$ can no longer be considered as an effectively massless field in de Sitter space. As a consequence, its fluctuations will be blocked. This is demonstrated in Figures~\ref{fig:FlatDirectionPDF} and~\ref{fig:FlatDirectionVariance}, which depicts the  case $\lambda = g_1 = g_2 =1$.

\FIGURE[ht]{
\epsfig{width=7.0cm,height=11.0cm,angle=-90, file=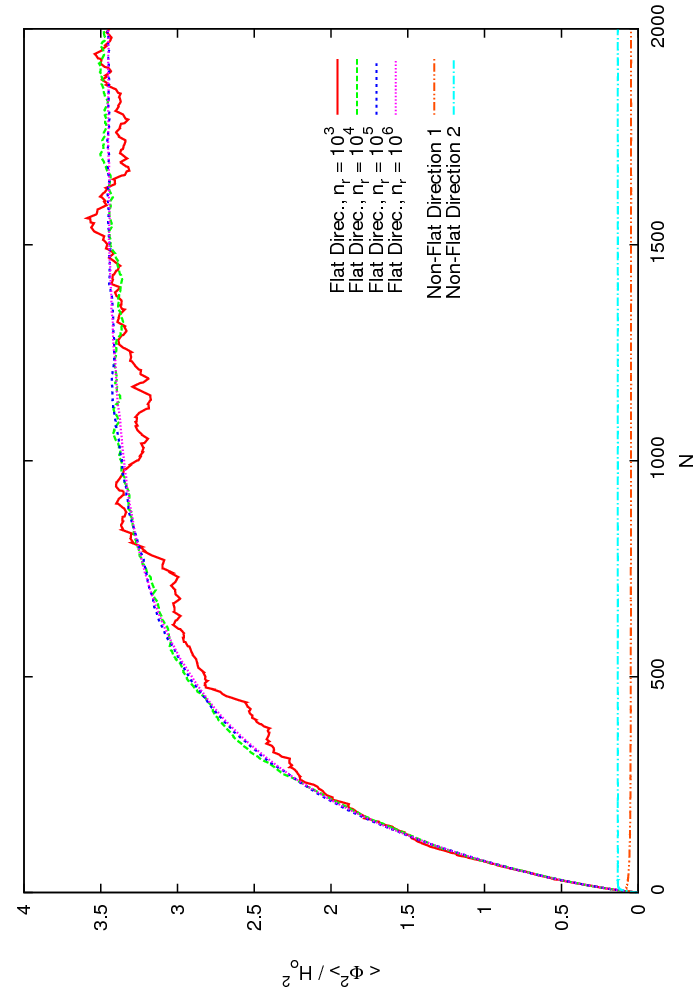}
\caption{Evolution in time of the variance of the flat direction $\phi$ in the strong coupling regime $\lambda = g_1 = g_2 =1$. Also shown is the evolution of the variance of the non-flat directions. The different plots of the flat direction correspond to different reconstructions of the evolution of $\left\langle \phi^2 \right\rangle$ versus $N$, as inferred from realizations with different number of independent runs, $n_r = 10^3~,10^4,~10^5$ and $10^6$. Even for the poorly sampled case of $n_r = 10^3$, the physics is already well captured: the growth of the flat-direction variance stops after~$\sim 1000$ efolds, reaching an asymptotic value of $\left\langle \phi^2 \right\rangle \approx 3.5 H_o^2$, much smaller than the expected value $N H_o^2/4\pi^2$ for a free flat direction massless field.}
\label{fig:FlatDirectionVariance}
}

In Figure~\ref{fig:FlatDirectionPDF} we show the evolution in time of the PDF of the flat directions. In Figure~\ref{fig:FlatDirectionVariance} we show the time evolution of the variance of each field. As can be clearly seen in the plots, it takes quite a long time, of the order of $\sim 1000$ e-folds, to reach a stationary regime. Once in this regime, the form of the flat direction PDF does not evolve in time anymore (red curves in Figure~\ref{fig:FlatDirectionPDF}), and the variance growth stops, reaching a final asymptotic amplitude (see Figure~\ref{fig:FlatDirectionVariance}). Note nevertheless, that during the first few tens of e-folds, the variance grows linearly with $N$, but still much more slowly than in the purely massless case. After the first hundred e-folds, the growth of the variance slows down and approaches the stationary regime. After $N \approx 1000$ efolds, the flat-direction finally enters into the stationary regime and its variance reaches an asymptotic constant amplitude $\left\langle \phi^2 \right\rangle \approx 3.5 H_o^2 $, much smaller than the expected amplitude for a free massless field $\left\langle \phi^2 \right\rangle = (N/4\pi^2) H_o^2$.

In all of this, the non-renormalizable term plays no role, since the flat direction field never manages to fluctuate further than few units of $H_o$. For $\lambda=1$, the amplitude of the flat direction in the stationary regime, is $\sim \mathcal{O}(\sqrt{M_p/H_o})$ orders of magnitude smaller than the value needed for the non-renormalizable term to become relevant.

However, all this depends on the magnitude of the couplings. In Figure~\ref{fig:VarianceChangingLambda}, we show the time evolution of the variance of the flat direction variance with different values of the Yukawa coupling $\lambda$, whilst $g_1 = g_2 = 1$ as before. For $\lambda < \mathcal{O}(10^{-2})$, the flat-direction
appears to fluctuate as one would expect from a massless field in de Sitter space. For $\lambda \sim \mathcal{O}(10^{-1})$, however, we already see some deviation from the purely massless case, since the flat-direction continues to grow for at least few thousand e-folds, albeit more slowly. For values of $\lambda$ between 0.1 and 1, we observe gradual blocking of the flat-direction fluctuations.

Thus, at the qualitative level, we may conclude that fluctuations within the F-term block the fluctuations along the flat directions that involve the third generation squarks and sleptons, while those involving only the first generation are likely to remain free to fluctuate. However, this is a conclusion based on the F-term alone and must be supplemented by the additional blocking
provided by the D-term, as will be discussed in the next subsection.

\FIGURE[ht]{
\epsfig{width=7.0cm,height=11.0cm,angle=-90, file=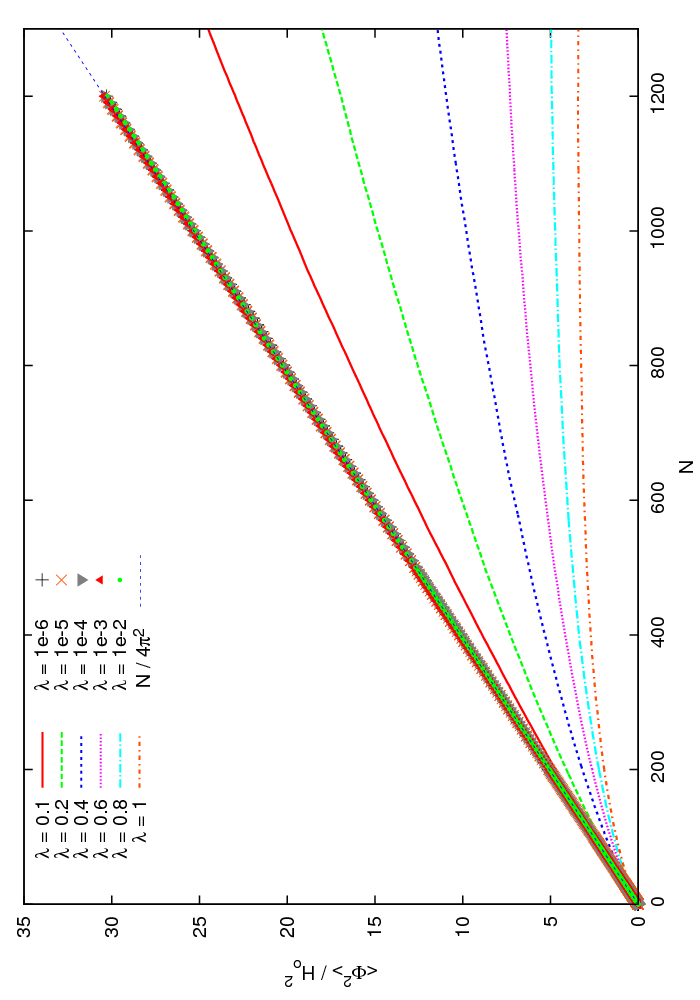}
\caption{Time evolution of Variance of the Flat direction, for different Yukawa couplings $\lambda$, from $\lambda  = 10^{-6}$ to $\lambda = 1$. For the smallest cases, $\lambda < \mathcal{O}(10^{-2})$, the flat-direction fluctuates as in the case of a massless free field in de Sitter space. For $\lambda \sim \mathcal{O}(10^{-1})$, the flat-direction
continues to grow for at least thousand e-folds, albeit more slowly than in the massless free case. For $\lambda \sim \mathcal{O}(0.1)-\mathcal{O}(1)$, the freedom of the flat-direction to fluctuate is gradually blocked as $\lambda$ increases.}
\label{fig:VarianceChangingLambda}
}

We should like to stress that the PDF properties do not depend on the richness of the statistics. As can be seen in Figure~\ref{fig:FlatDirectionVariance}, $n_r = 10^3$ runs capture already all the essentials. There we show the evolution of the variance also for $n_r = 10^4,~10^5$ or $10^6$ independent runs, and all plots show the same behaviour, the only difference being that the richer the statistical sampling (i.e. the bigger $n_r$) the less noisy the plots are. As yet another check (not shown in the plots), we have compared the outputs from the same number of independent runs but different step size $dN$. We have found that if the simulation incorporates more than $100$ stochastic kicks per e-fold (i.e.~if $dN < 0.01$), the properties of the PDF remain the same as compared to the $dN = 0.01$ case, no matter how small we make $dN$. For simulations with $n_r = 10^5$ or more independent runs, we cannot even detect any difference between choosing $dN = 0.01$ or $dN = 0.001$.

\FIGURE[ht]{
\epsfig{width=7.0cm,height=11.0cm,angle=-90,file=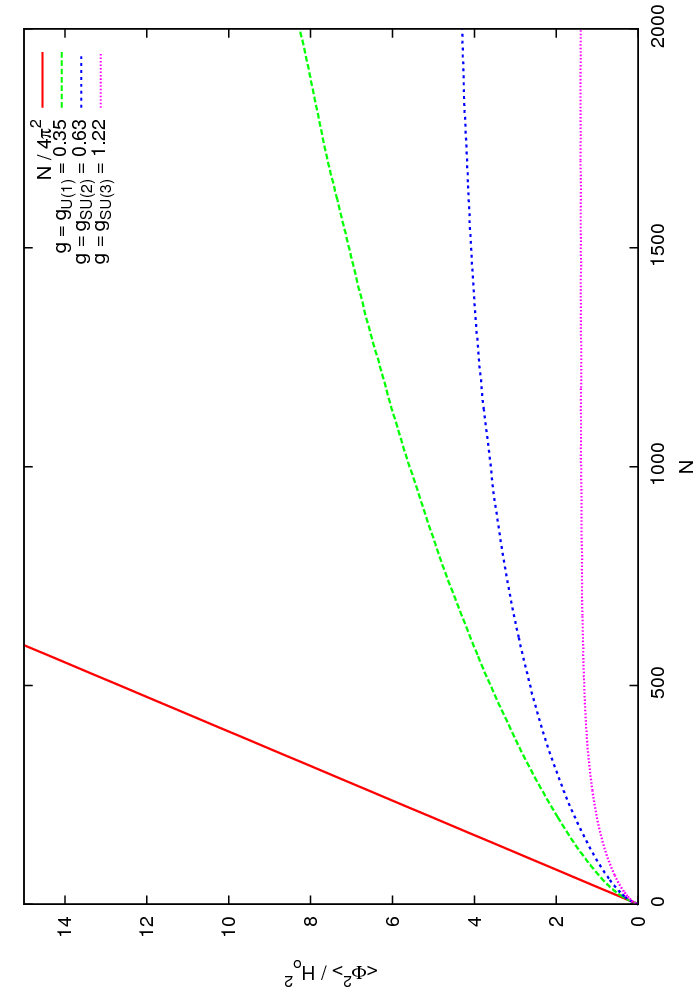}
\caption{Time evolution of the variance of the field $L$ ($H$ is practically identical), for different gauge couplings $g$ in the potential~(\ref{dtermcase}). The larger the coupling is, the stronger the blocking of the flat direction fluctuations, and thus, the smaller the amplitude reached. With relatively strong $SU(3)$ interactions, the fields reach a stationary regime after only $N \sim 200-300$ efolds. In the $SU(2)$ case, the fields are just entering into the stationary phase after the first $\sim 1000-2000$ efolds, whereas in the $U(1)$ case, the amplitude of the fields is still growing after $N = 2000$ efolds, albeit with a much smaller amplitude than in the free massless case (straight line in the Figure).}
\label{fig:DpotentialVariance}
}
%%%%%%%%%%%%%%%%%%%%%%%%%%%%%%%%%%%%%%%%%%%%%%%%%%%%%%%%%%%%%%%%%%%%%%%%%%%%%%%%%%%%%%%%%%%%%%%%%%%%%%%%%%%%
%%%%%%%%%%%%%%%%%%%%%%%%%%%%%%%%%%%%%%%%%%%%%%%%%%%%%%%%%%%%%%%%%%%%%%%%%%%%%%%%%%%%%%%%%%%%%%%%%%%%%%%%%%%%
\subsection{Breaking D-flatness}

Let us now focus on fluctuations within the D-terms. We set the F-term (i.e. the Yukawas) to zero and adopt the potential~(\ref{dtermcase}). We have thus two fields, $H$ and $L$, interacting with the coupling $g$, for which the configurations $H = \pm L$ represent a flat direction.

\FIGURE[ht]{
\epsfig{width=4.0cm,height=4.0cm,angle=0,file=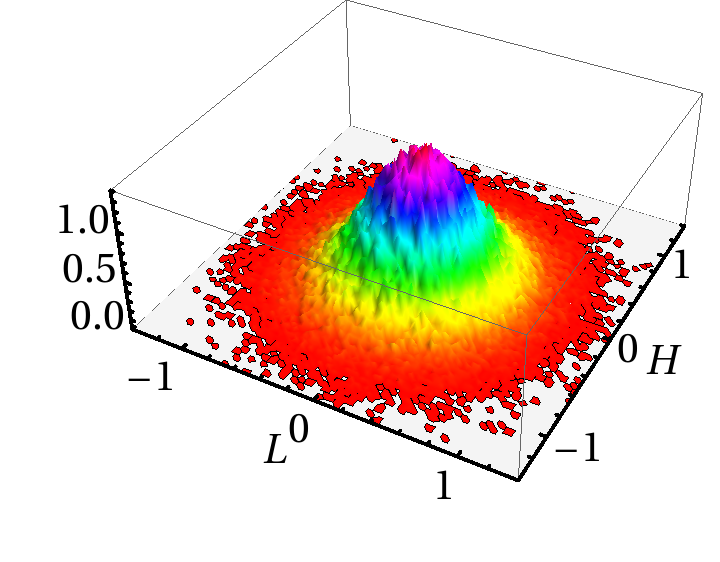}\hspace{0.5cm}\epsfig{width=4.0cm,height=4.0cm,angle=0,file=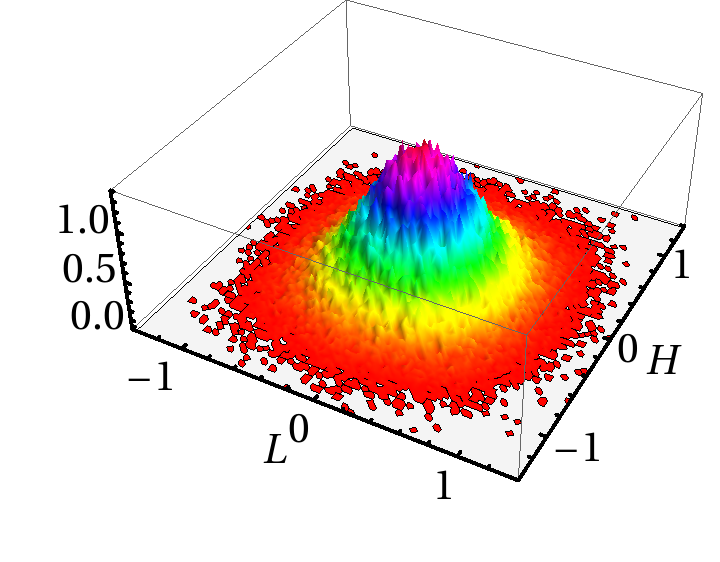}\hspace{0.5cm} \epsfig{width=4.0cm,height=4.0cm,angle=0,file=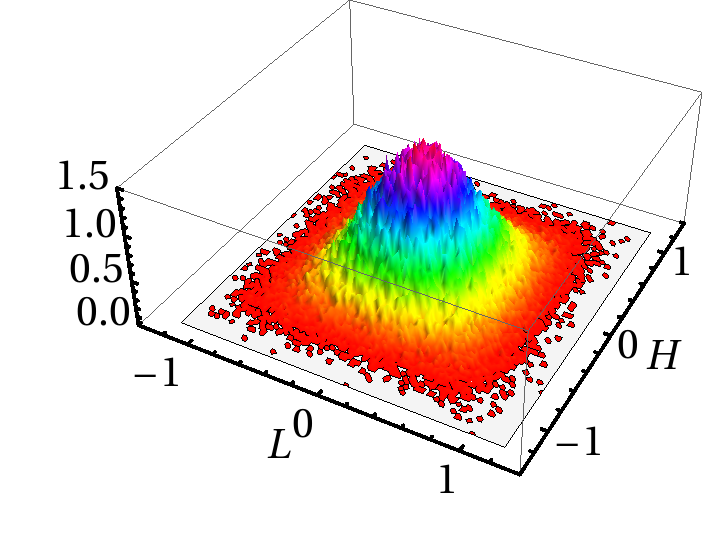}\\
\epsfig{width=4.0cm,height=4.0cm,angle=0,file=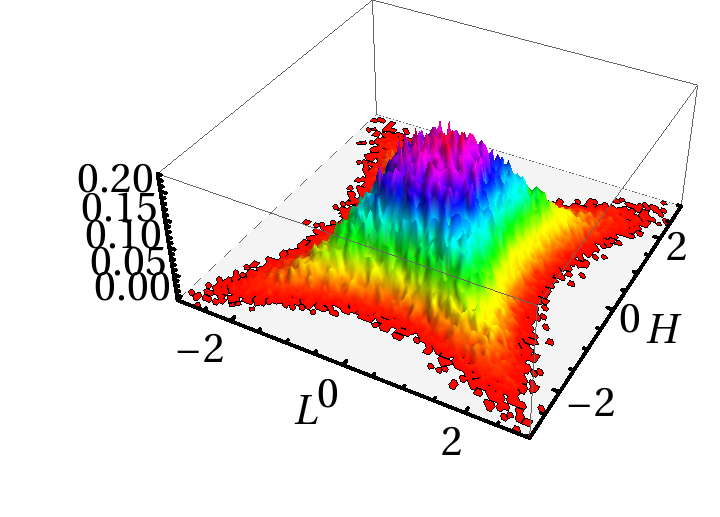}\hspace{0.5cm}\epsfig{width=4.0cm,height=4.0cm,angle=0,file=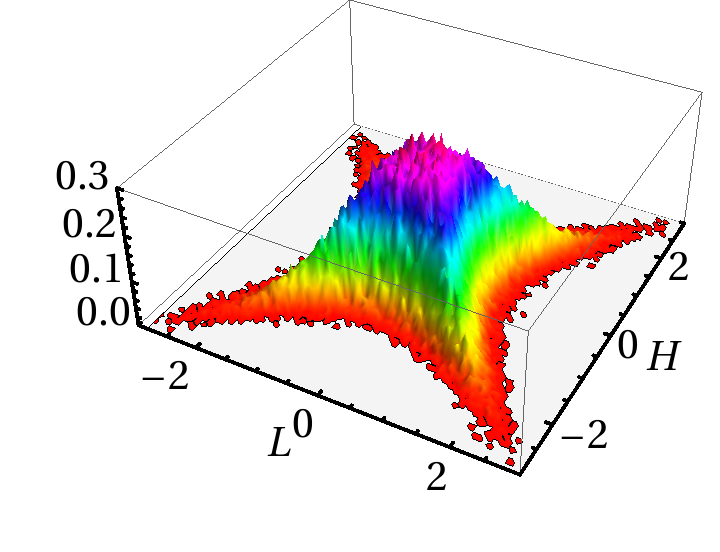}\hspace{0.5cm} \epsfig{width=4.0cm,height=4.0cm,angle=0,file=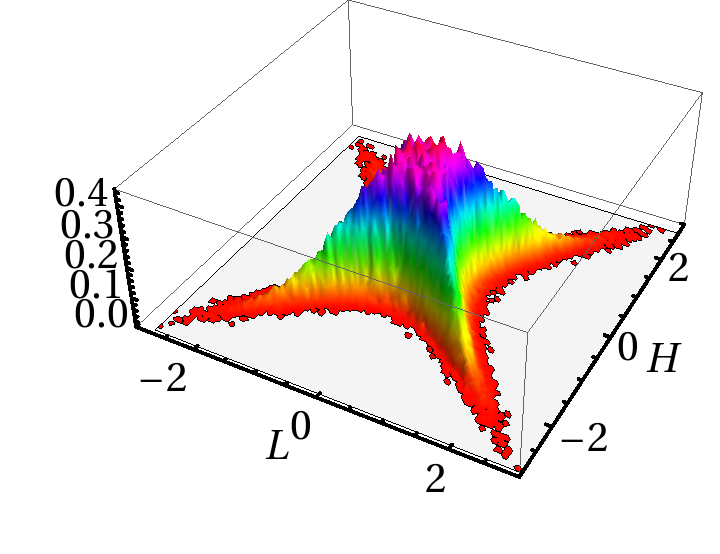}\\
\epsfig{width=4.0cm,height=4.0cm,angle=0,file=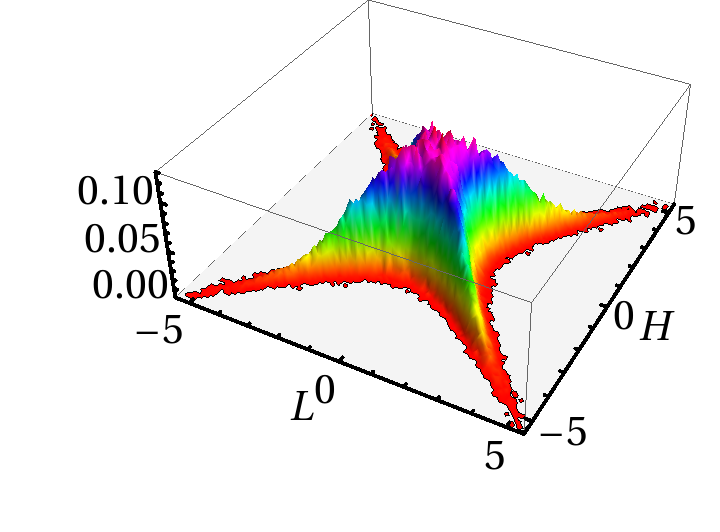}\hspace{0.5cm} \epsfig{width=4.0cm,height=4.0cm,angle=0,file=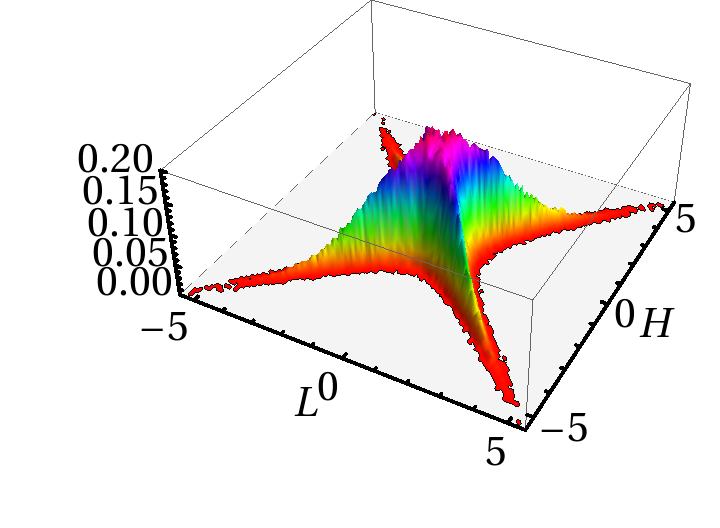}\hspace{0.5cm} \epsfig{width=4.0cm,height=4.0cm,angle=0,file=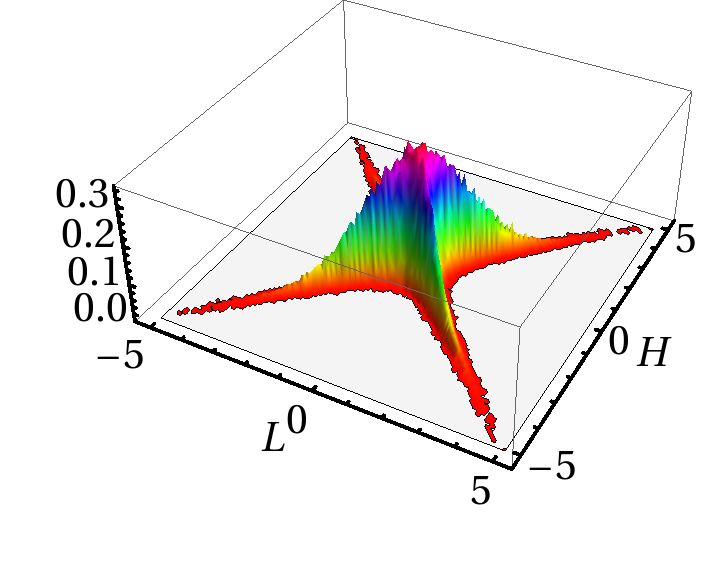}\\
\epsfig{width=4.0cm,height=4.0cm,angle=0,file=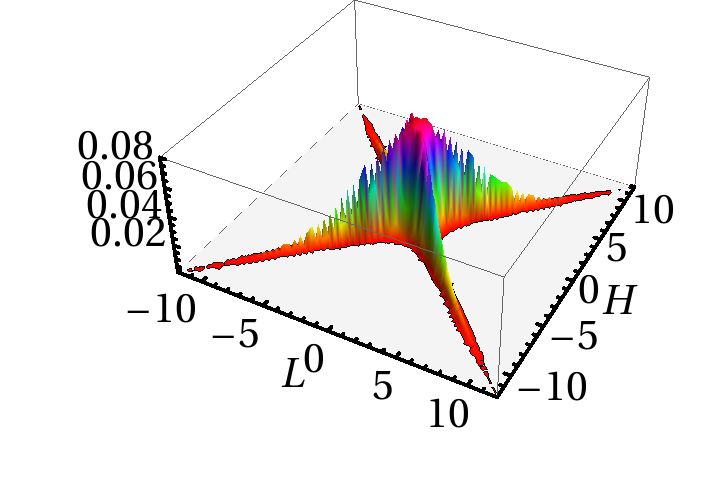}\hspace{0.5cm} \epsfig{width=4.0cm,height=4.0cm,angle=0,file=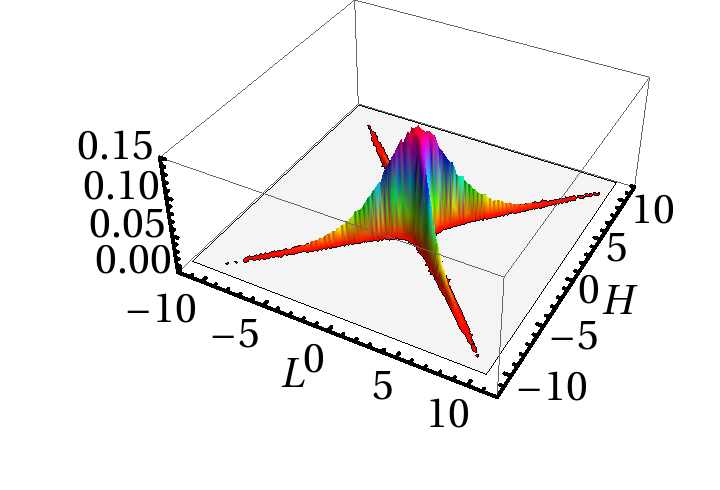}\hspace{0.5cm} \epsfig{width=4.0cm,height=4.0cm,angle=0,file=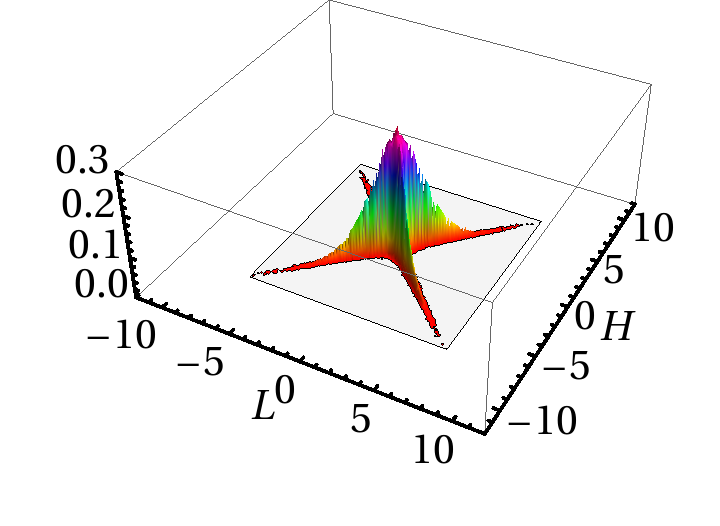}\\
\caption{Sequence of snapshots at different moments of the evolution of the 2-dimensional PDF $\mathcal{P}(L,H)$. From left to right we consider the case of having a $U(1)_Y$, $SU(2)_L$ and $SU(3)_c$ gauge coupling. From top to bottom, the rows correspond to $N = 5, 40, 150$ and $1000$ efolds (note the change of scale in the field space for different efolds). Looking at the snapshots in the bottom row, one can appreciate how little the flat direction fields
fluctuate at large couplings.}
\label{fig:2dHistograms}
}

In Figure~\ref{fig:DpotentialVariance} we show the evolution of the variance of $L$ and $H$ for different strengths of the coupling between them. As in the case with an F-flat direction, we see again a gradual blocking in time of the freedom of these fields to fluctuate. The fluctuations of $H$ and $L$ are restricted more severely the stronger the gauge coupling. This can be observed in Figure~\ref{fig:DpotentialVariance}, which shows the situation for the gauge couplings\footnote{To be more precise, one should let  the gauge couplings run to the scale of inflation $\sim \sqrt{HM_p}$. However, for our purposes, this would not change -- qualitatively speaking -- anything; the $U(1)$ and $SU(2)$ couplings would simply grow slightly, and thus the flat direction fluctuations in that case would result slightly more blocked, whereas although the $SU(3)$ coupling would indeed decrease, it would never become small enough for the flat direction to fluctuate as if it were a free massless field. } of the (MS)SM, $g_1 = 0.35$, $g_2 = 0.62$ and $g_3 = 1.23$, for $U(1)_Y$, $SU(2)_L$ and $SU(3)_c$ interactions, respectively.

Of course the variance of $L$ or $H$ is not exactly the same as the variance of the flat direction, but still it gives a reliable measure about the freedom of the flat directions to fluctuate. To completely characterize the statistical properties of the flat directions, we should rather obtain instead the 2-dimensional PDF $\mathcal{P}(L,H)$, which describes the simultaneous distribution of both $H$ and $L$. In particular, $\mathcal{P}(L',H')dLdH$ represents the probability of finding $(L,H)$ within the field-space region $[L',L'+dL]\times[H',H'+dH]$. Therefore, similarly to the one-dimensional case, the $\mathcal{P}(L,H)$ function can be obtained from the numerics, simply by dividing the 2-dimensional unit-normalized histogrammatic distribution of the fields, at each time by the field bin steps $\Delta H$ and $\Delta L$.

In Figure~\ref{fig:2dHistograms} we show the evolution of the field fluctuations in the $(L,H)$ plane for all the (MS)SM gauge couplings, by plotting $\mathcal{P}(L,H)$ at different moments. One can clearly see the probability leaking into the flat directions $L = \pm H$. As expected from the behaviour of the variances shown in Figure~\ref{fig:DpotentialVariance}, such leaking probability is of course higher the weaker the coupling. This is visualized in Figure~\ref{fig:2dHistograms}, where we have fixed the scale in the horizontal to be same in all the snapshots taken at a given  time for each different couplings. One can very clearly appreciate that the 'branches' developing in $\mathcal{P}(L,H)$ along the flat directions, reach greater distances in the $(L,H)$ plane the weaker the gauge coupling is. For the $U(1)_Y$ case, the weakest of the (MS)SM couplings, the flat direction branches reach values as high as $12 H_o$. But even in this case, in which the flat directions are the least constrained to fluctuate (compared to the $SU(2)_L$ and $SU(3)_c$ cases), such high amplitude regions are still highly unlikely. The 2-dim PDF obtained for the different coupling regimens are very suppressed for those high field values, if we compare them with the PDF for the case of a massless flat direction.

To assess the degree of blockage of the flat directions in the different coupling regimes, we show in Table 1 the probability of the fields to be above certain fixed thresholds. We then compare with the probability that one would expect for a massless non-interacting flat direction field. The numbers speak for themselves.

\begin{center}
\begin{table}
\hspace{2cm}
\begin{tabular}{||l | c c c ||}
\hline
\hspace{2cm}Prob $\backslash$ $N$ & $N = 40$ & $N = 150$ & $N = 1000$\\
\hline
Prob($|H \pm L| > 2; g \ll 1$) & 14 \% & 59 \% & 92\% \\
Prob($|H \pm L| > 2; g_{U(1)}$)  & 0.30 \% & 8.2 \%  & 29 \%\\
Prob($|H \pm L| > 2; g_{SU(2)}$) & 0.27 \% & 6.6 \%  & 21 \%\\
Prob($|H \pm L| > 2; g_{SU(3)}$) & 0.16 \% & 3.8 \%  & 9 \% \\
\hline
Prob($|H \pm L| > 5; g \ll 1$)  & 0.04 \% & 12 \% & 73 \%\\
Prob($|H \pm L| > 5; g_{U(1)}$)  & 0.009 \% & 0.56 \%  & 12 \%\\
Prob($|H \pm L| > 5; g_{SU(2)}$) & 0.0010 \% & 0.35 \%  & 7.4 \%\\
Prob($|H \pm L| > 5; g_{SU(3)}$) & 0.0007 \% & 0.11 \%  & 1.1 \%\\
\hline
\end{tabular}
\caption{Here Prob($|H \pm L| > A; g_c$) represents the probability of finding any flat direction with an amplitude with absolute value greater than $A H_o$, when the coupling is $g_c$. The case $g \ll 1$ represent the free field case.}
\label{tab:a}
\end{table}
\end{center}

%%%%%%%%%%%%%%%%%%%%%%%%%%%%%%%%%%%%%%%%%%%%%%%%%%%%%%%%%%%%%%%%%%%%%%%%%%%%%%%%%%%%%%%%%%%%%%%%%%%%%%%%%%%%
%%%%%%%%%%%%%%%%%%%%%%%%%%%%%%%%%%%%%%%%%%%%%%%%%%%%%%%%%%%%%%%%%%%%%%%%%%%%%%%%%%%%%%%%%%%%%%%%%%%%%%%%%%%%

\section{Conclusions}
\label{sec:Conclusions}

During inflation, all the scalar fields fluctuate. The non-flat directions are coupled with each other either by virtue of Yukawa interactions or, as is often the case, via the D-term. Their variances will spread out to values of the order of a fraction of $H_o$ in a matter of relatively few efolds. These variances act as effective mass terms for the flat direction $\phi$, whose fluctuations therefore become blocked and eventually saturate as equilibrium is reached with $\langle\phi^2\rangle = \alpha H_o^2, \alpha \sim \mathcal{O}(1)-\mathcal{O}(10)$, unless the coupling is very weak. The stronger the coupling strength, the more effective is the blocking, as is evident in Figures~\ref{fig:VarianceChangingLambda} and~\ref{fig:DpotentialVariance}.

We have verified these expectations quantitatively in Sect. \ref{sec:StochasticDynamics}, where we set up the the coupled system of Langevin equations, which account for the field fluctuations during inflation, and studied the fluctuations numerically. For simplicity, we focused on the F-terms and D-terms separately. Our main results are encoded in Figures~\ref{fig:VarianceChangingLambda} and~\ref{fig:DpotentialVariance}, which elucidate the spreading of the field variances as the function of the number of efolds for different coupling strengths.

We have assumed that initially, at the beginning of inflation, all the fields are at the origin of the potential. Any random displacements around
the origin are not likely to change qualitatively our conclusions, except of course for those initial conditions where the flat direction
field has from the outset an amplitude much larger than $H_o$. It should also be emphasized that we assume inflation in the background with all the
susy scalars acting as spectators. This is not the case in MSSM inflation \cite{MSSinf}, which makes use of a saddle point along the flat direction.
Thus, the present considerations do not apply to MSSM inflation, where the initial amplitude is put in by hand.

However, many other cosmological deliberations that tacitly assume a large amplitude along the flat direction generated during inflation should be reconsidered. There are two issues one should bear in mind when discussing flat directions and inflation: the blocking of fluctuations due to non-flat directions, discussed in the present paper, and the number of efolds actually needed to reach equilibrium, which concerns the inflationary sector rather than the more precisely defined susy scalars.

Finally, although we have considered F-flatness and D-flatness separately, it might be possible to solve the Langevin equations for the full MSSM scalar field contents and follow the fluctuations of all the fields, including the simultaneous fluctuations along many flat directions. This would be an interesting exercise.

\acknowledgments
We thank Anupam Mazumdar for numerous discussions, and Marco Panero for his help with numerical random generators. We also acknowledge the authors of the publicly available distribution ``Mersenne Twister''~\cite{Random Generator}, which we used for generating the stochastic noises in the Langevin equations. KE and DGF are respectively supported by the Academy of Finland grants 131454 and 218322. GR is supported by the Gottfried Wilhelm Leibniz programme of the Deutsche Forschungsgemeinschaft.

%%%%%%%%%%%%%%%%%%%%%%%%%%%%%%%%%%%%%%%%%%%%%%%%%%%%%%%%%%%%%%%%%%%%%%%%%%%%%%%%%%%%%%%%%%%%%%%%%%%%%%%%%%%%
%%%%%%%%%%%%%%%%%%%%%%%%%%%%%%%%%%%%%%%%%%%%%%%%%%%%%%%%%%%%%%%%%%%%%%%%%%%%%%%%%%%%%%%%%%%%%%%%%%%%%%%%%%%%
\section*{Appendix. Some aspects of the derivation of the Langevin equations}
\label{sec:Appendix}

Let us consider some physical scale $L$ and IR/UV decompose any field $\phi$ with respect to the physical volume $V_L \sim  1/L^3$. Choosing our frame of reference as centered in such a region, the IR part $\phi_{\IR}$ is coordinate-independent within $V_L$, and it can be identified in fact with the volume average of $\phi$ over $V_L$, i.e. \beq
\phi_\IR \sim \langle \phi(\bx,t)\rangle_{_{V_L}}\equiv \frac{1}{V_{L}}\int_{V_L} d^3\bx\,\,\phi_i\,.
\eeq
This holds as long as the window function $W$ used in the IR/UV decomposition has compact support in real space. In $de~Sitter$ space, the natural choice is $L = H_o^{-1}$. Considering a region of physical volume $V_{H_o} \sim H_o^{-3}$, each field can then be decomposed as $\phi_i(\bx,t) = \Phi_i(t) + \varphi_i(\bx,t)\,,$ with $\Phi_i$ and $\phi_i$ obtained according to eqs.~(\ref{eq:IR/UV})-(\ref{eq:IR/UVbis}), as long as $|\bx| \ll H_o^{-1}$. The IR part $\Phi_i$ can be indeed interpreted as the contribution to $\phi_i$ from all field wavelengths larger than $H_o^{-1}$ at the given region.

Let us now derive the Langevin equations in some detail. First of all, note that the IR/UV parts of the derivative of the field are not the same\footnote{If we consider, however, the derivative of the total $\phi$, and not the derivative of each of its IR and UV parts separately, then it is true that $\dot\phi(\bx,t)$ = $(\phi_\IR)^\cdot(\bx,t) + (\phi_\UV)^\cdot(\bx,t)$ = $\dot\phi_\IR(\bx,t) + \dot\phi_\UV(\bx,t)$.} as the derivative of the IR/UV parts of the field, i.e. $({\rm IR}\lbrace\phi\rbrace)^\cdot(\bx,t) \neq$ $ {\rm IR}\lbrace\dot\phi\rbrace$ and $({\rm UV}\lbrace\phi\rbrace)^\cdot(\bx,t)$ $\neq {\rm UV}\lbrace\dot\phi\rbrace$. This curious property, which emerges simply from the fact that in an expanding universe the window function $W(\bk,t)$ for a fixed physical volume depends on time, has nonetheless profound consequences for the dynamics of fields living in De Sitter space. This property is indeed the key factor behind the derivation of the Langevin eqs.~describing the stochastic behaviour of the IR $dof$.

In the hamiltonian picture the $eom$ are
\begin{eqnarray}\label{eq:HamiltonianEqs2}
\dot\phi_i = \pi_i\,,\hI\hV \dot\pi_i + 3H_o\pi_i &=& \frac{1}{a^2}\nabla^2\phi_i - d_iV\,,
\end{eqnarray}
with $d_iV \equiv \partial V/\partial\phi_i$. We should thus IR/UV decompose independently $\phi_i$ and $\pi_i$ as $\phi_i(\bx,\eta) = \Phi_i(\eta) + \varphi_i(\bx,\eta)$ and $\pi_i(\bx,\eta) = \Pi_i(\eta) + \delta\pi_i(\bx,\eta)$, and introduce such decomposition into eqs.~(\ref{eq:HamiltonianEqs2}). Assuming that $V(\lbrace \phi_j\rbrace)$ is infinitely differentiable with respect to any field $\phi_i$, then \begin{eqnarray}\label{eq:MultiExpansion}
d_iV(\lbrace \phi_j\rbrace) = D_{i}\bar V(\lbrace \Phi_j\rbrace) + \sum_j (D_{ij}\bar V)\varphi_j + \cdots\,,
\end{eqnarray}
with $D_i = \partial/\partial\Phi_i$, $D_{ij} = \partial^2/\partial\Phi_i\partial\Phi_j$, and $\bar V \equiv V(\lb \Phi_j\rb)$. We should then plug this expression for $d_iV$ into the eqs.~(\ref{eq:HamiltonianEqs2}), and separate the evolution of the IR $dof$ from the UV $dof$. One finds this way
\begin{eqnarray}\label{eq:IR/UV hamiltonian}
&& \dot\Phi_i + \dot\varphi_i - \Pi_i - \delta\pi_i = + \int{\frac{d^3\bk}{\ps}\,e^{-i\bk\bx}\,\phi_i(\bk,t)\,\dot W(k,t)}\\
\label{eq:IR/UV hamiltonian2}
&& \dot\Pi_i + \dot\delta\pi_i + 3H_o(\Pi_i+\delta\pi_i) - \frac{1}{a^2}\nabla^2\varphi_i + D_i\bar V + \sum_j (D_{ij}\bar V)\varphi_j + ... = \int{\frac{d^3\bk}{\ps}\,e^{-i\bk\bx}\,\pi_i(\bk,t)\,\dot W(k,t)}\,,\nn\\
\end{eqnarray}
where we have neglected $\nabla^2\Phi_i$, since this term is sub-dominant in determining the evolution of the average value of the field $\Phi_i \sim \langle \phi \rangle_{V_L}$ inside a volume $V_L \sim H_o^{-3}$.
 %since we are considering length scales inside the volume , and thus  is coordinate-independent. Note that if the fields had some spatial features at distances much greater that the characteristic length of the volume considered, then $\Phi_i$ components might really dependend on the position $\bx$, though understanding such coordinates as labeling points separated a greater distance than $L \sim 1/H$. When these long-range features are absent, then the fields is said to be in the \textit{ultra-local} limit, and gradients of the IR components might be neglected (versus the time derivatives). However, if the gradients over scales longer than $\sim H^{-1}$ cannot be ignored, one should plug in back the term $-\nabla^2\Phi_i$ into the above equations, in order to describe correctly the evolution of the IR part of the fields over big scales.
Ignoring short wavelength interactions, the UV $dof$s satisfy
\begin{eqnarray}
 \dot\varphi_i - \delta\pi_i = 0\hspace*{4cm}\\
 \dot{\delta\pi}_i + 3H_o\delta\pi_i - \frac{1}{a^2}\nabla^2\varphi_i + D_i\bar V + \sum_j (D_{ij}\bar V)\varphi_j + \cdots = 0\,,
\end{eqnarray}
so, subtracting these eqs.~to eqs.~(\ref{eq:IR/UV hamiltonian})-(\ref{eq:IR/UV hamiltonian2}), we are then left with the equations for the IR $dof$ as
\begin{eqnarray}\label{eq:LangevinEqs}
&& \dot\Phi_i = \Pi_i + \int{\frac{d^3\bk}{\ps}\,e^{-i\bk\bx}\,\phi_i(\bk,t)\,\dot W(k,t)}\\
\label{eq:LangevinEqsBis}
&& \dot\Pi_i = 3H_o\Pi_i + D_i\bar V + \int{\frac{d^3\bk}{\ps}\,e^{-i\bk\bx}\,\pi_i(\bk,t)\,\dot W(k,t)}
\end{eqnarray}

Considering the previous equations as operator equations, we arrive at the usual interpretation of the eq. of the IR modes as Langevin eqs (see main text). To completely characterize these equations, we need however to choose a window function. The simplest choice for it is a step function
\begin{equation}
 W(k,t) = \theta(\epsilon\, aH_o-k) = 1 - \theta(k-\epsilon\,aH_o)\,,
\end{equation}
where the number $\epsilon<1$. Although this window function does not have a compact support in real space it suffices for our purposes as we will not be interested in the exact spatial correlations of $\Phi_i$. In such a case $\dot W(k,t) = \epsilon a(t)H_o^2\delta(\epsilon aH_o-k)$. Using this and the usual normalization of the creation/annihilation operators,
\begin{eqnarray}
\langle \hat a_i(\bq)\hat a_j^\dagger(\bq')\rangle = \ps \delta_{ij} \delta^3(\bq-\bq')\,,\hspace*{1.2cm}\\
\langle \hat a_i(\bq)\hat a_j(\bq')\rangle = \langle \hat a_i^\dagger(\bq)\hat a_j^\dagger(\bq')\rangle = \langle \hat a_i^\dagger(\bq)\hat a_j(\bq')\rangle = 0\,,
\end{eqnarray}
one arrives at
\begin{eqnarray}
&& s^{(\phi)}_{ij}(\bx,\bx',t,t') = \frac{\epsilon^3H_o^4}{2\pi^2}a^3(t)|f_{\phi}(q,t)|^2_{_{q = \epsilon aH_o}}j_0(aH|\bx-\bx'|)\delta(t-t')\,,\\
&& s^{(\pi)}_{ij}(\bx,\bx',t,t') = \frac{\epsilon^3H_o^4}{2\pi^2}a^3(t)|f_\pi(q,t)|^2_{_{q = \epsilon aH_o}}j_0(aH|\bx-\bx'|)\delta(t-t')\,,
\end{eqnarray}
where $j_0(x)$ is a spherical Bessel function of order 0. Had we chosen another function for $W(k,t)$, like a gaussian or some other function smoothed around the scale $\epsilon\,aH$, then correspondingly we would have found, instead of a 'sharp function' $\delta(t-t')$, rather a smooth function peaked at $t = t'$. Moreover, note a couple of things. First, we only need $s^{(\phi)}_{ij}$ and $s^{(\pi)}_{ij}$ at $\bx=\bx'$, since we are only interested in the random kicks that an IR mode $\Phi_i$ receives in its domain of size $\sim H_o^{-3}$, and not in the correlation between kicks at different (causally disconnected) domains. Therefore, the spherical Bessel functions will be dropped, since $j_0(0) = 1$. Secondly, these correlators are indeed divergent at equal times $t = t'$, as a reflection of the fact that we have chosen a 'sharp' step-function for $W(k,t)$. However, the presence of $\delta(t-t')$ will not be a problem for our purposes, since as explained in Section III, the correlators we really need to compute are
\begin{eqnarray}
&& S^{(\phi)}_{ij}(t,dt) \equiv \int_{t}^{t+dt}\int_{t'}^{t'+dt}s^{(\phi)}_{ij}(\tilde t,\tilde t')d\tilde td\tilde t' = \epsilon^3\delta_{ij}\frac{H_o^4}{2\pi^2}\int_{t}^{t+dt}d\tilde t\,a^3(\tilde t)|f_{\phi}(q,\tilde t)|^2_{_{q = \epsilon aH_o}},\nn\\\\
&& S^{(\pi)}_{ij}(t,dt) \equiv \int_{t}^{t+dt}\int_{t'}^{t'+dt}s^{(\pi)}_{ij}(\tilde t,\tilde t')d\tilde td\tilde t' = \epsilon^3\delta_{ij}\frac{H_o^4}{2\pi^2}\int_{t}^{t+dt}d\tilde t\,a^3(\tilde t)|f_{\pi}(q,\tilde t)|^2_{_{q = \epsilon\,aH_o}},\nn\\
\end{eqnarray}

In a pure $de\,\,Sitter$ background for Inflation, the solution to the mode equations (with boundary conditions matching Minkowski modes at $k \rightarrow \infty$) are~\cite{LindeBook}
\begin{eqnarray}\label{eq:deSitterModeFunctions}
f_{\phi}(\bk,t) = \frac{\sqrt{\pi}}{2}\frac{H_o}{k^{3/2}}\left(x^{3/2}H_{3/2}^{(2)}(x)\right)\,,\hV
f_{\pi}(\bk,t) = \frac{\sqrt{\pi}}{2}\frac{H_o}{k^{1/2}a(t)}\frac{d\left(x^{3/2}H_{3/2}^{(2)}(x)\right)}{dx}
\end{eqnarray}
with $H_{3/2}^{(2)}(x)$ a second-kind Hankel function of order $\frac{3}{2}$, $x \equiv k\eta$ and $\eta = \int dt/a(t)$ the usual conformal time. Using these mode functions evaluated at $k = \epsilon aH_o$, one can therefore obtain the exact form of the correlators $S^{(\phi)}_{ij}$ and $S^{(\pi)}_{ij}$. In the limit in which $dt \ll 1/H_o$, one finds
\begin{eqnarray}\label{eq:CorrelatorsNaturalVariables}
S^{(\phi)}_{ij}(t,dt) = (1+\epsilon^3)\delta_{ij}\frac{H_o^3}{4\pi^2}\,dt\,,\hI
S^{(\pi)}_{ij}(t,dt) = \epsilon^4\delta_{ij}\frac{H_o^5}{(2\pi)^2}\,dt\,,
\end{eqnarray}
We have used these noise correlators evaluated at $\epsilon = 0.1$, and observed that the results do not change if one makes $\epsilon$ even smaller.

\end{document}